\documentclass[epjST,final]{svjour}
\usepackage{graphicx,amsfonts,amsmath,amssymb,xcolor,subfigure}
\usepackage[english]{babel}
\usepackage[numbers,sort&compress]{natbib}
\usepackage[latin1]{inputenc}

\newcommand{\pt}[2]{\frac{\mathrm{d}\, #1}{\mathrm{d}\, #2}\,}
\newcommand{\e}{\varepsilon}

\newcommand{\Z}{\mathbb{Z}}

\newcommand{\V}[1]{\mathbf{#1}}

\newcommand{\intl}[3]{\int\limits_{#1}^{#2}\mathrm{d}#3\,}
\newcommand{\bracket}[1]{\left(#1\right)}

\newcommand{\grad}{\mbox{\boldmath$\nabla$}}
\newcommand{\grb}[1]{\mbox{\boldmath $#1$}}

\newcommand{\eq}[1]{$\mathrm{Eq.}$~\eqref{#1}}
\newcommand{\eqs}[1]{$\mathrm{Eqs.}$~\eqref{#1}}
\newcommand{\figref}[1]{$\mathrm{Fig.}$~\ref{#1}}
\newcommand{\secref}[1]{$\mathrm{Sec.}$~\ref{#1}}

\newcommand{\av}[1]{\left\langle #1 \right\rangle}

\begin{document}

\title{Giant enhancement of hydrodynamically enforced entropic trapping in thin channels}

\author{S. Martens\inst{1}\fnmsep\thanks{\email{steffen.martens@tu-berlin.de}}
\and A. V. Straube\inst{2}
\and G. Schmid\inst{3}
\and L. Schimansky-Geier\inst{2}
\and P. H\"anggi\inst{3,4}}
\institute{
Department of Physics, Hardenbergstra\ss e 36, EW 7-1, Technische Universit\"{a}t Berlin, 10623 Berlin, Germany
\and Department of Physics, Humboldt-Universit\"{a}t zu Berlin, Newtonstr. 15, 12489 Berlin, Germany
\and Department of Physics, Universit\"{a}t Augsburg, Universit\"{a}tsstr. 1, 86135 Augsburg, Germany
\and Nanosystems Initiative Munich, Schellingstr. 4, D-80799 M\"unchen,
Germany}

\abstract{
Using our generalized Fick-Jacobs approach \cite{Martens2013,Martens2013b} and extensive Brownian dynamics simulations, we study particle transport through three-dimensional periodic channels of different height. Directed motion is caused by the interplay of constant bias acting along the channel axis and a pressure-driven flow. The tremendous change of the flow profile shape in channel direction with the channel height is reflected in a crucial dependence of the mean particle velocity and the effective diffusion coefficient on the channel height. In particular, we observe a giant suppression of the effective diffusivity in thin channels; four orders of magnitude compared to the bulk value.
}

\maketitle

\section{Introduction}

Effective control of mass and charge transport at microscale level is in the limelight of widespread timely activities in different contexts. Such endeavors involve \textit{Lab-on-chip} techniques \cite{Dietrich2006}, molecular sieves \cite{Keil2000,Berezhkovskii2005}, biological \cite{Berezhkovskii2008} and designed nanopores \cite{Pedone2010}, chromatography or, more generally, separation techniques of size-dispersed particles on micro- or even nanoscales \cite{Voldman2006}, to name but a few. Particle separation techniques use the fact that the particles' response to external stimuli, such as gradients
or fields, depends on their physical properties like surface charges, magnetization, size or shape \cite{HanggiRMP2009,Reguera2012}. Accordingly,
conventional methods for filtering particles involve centrifugal fractionation \cite{Harrison2002}, phoretic forces \cite{Dorfman2010,Pagonabarraga2011}
or external fields \cite{Macdonald2003}. Recently, novel devices for particle separation were proposed
\cite{HanggiRMP2009,Reguera2012,Bogunovic2012,Kettner2000,Zheng2013}. The capability of such devices for separation of particles is rooted in the effect of {\itshape entropic rectification}, i.e., the rectification of motion caused by broken spatial symmetry caused by asymmetric variations of the accessible local volume \cite{Zitserman2011,Dagdug2012,Motz2014}. The transport in channels with periodically varying cross-section exhibits peculiar transport phenomena \cite{Reguera2006,Burada2009_CPC,Rubi2010,Martens2012,Keyer2014,Keyer2014arxiv} which can be treated by means of the so-termed Fick-Jacobs formalism and its generalizations \cite{Martens2013,Jacobs,Zwanzig1992,Reguera2001,Kalinay2006,Kalinay2011,Martens2011,Martens2011b,Martens2012b,Ghosh2012PRE,Ghosh2012EPL}.

\begin{figure}[t]
  \centering
  \includegraphics[width=0.7\linewidth]{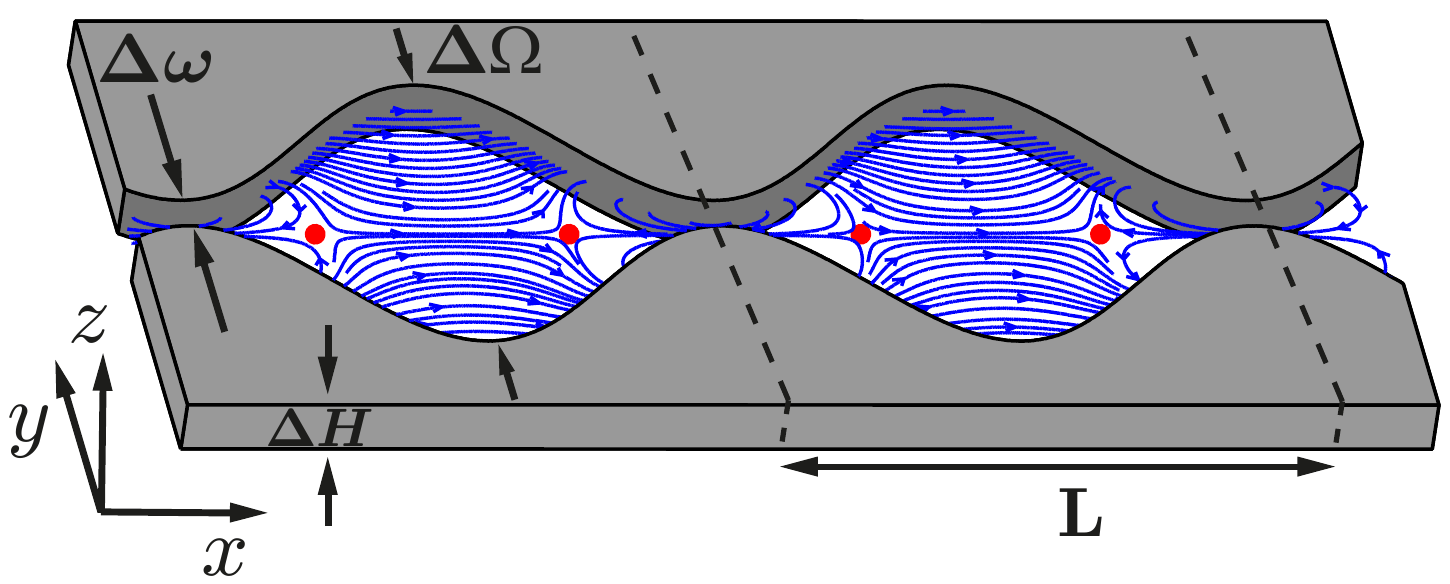}
  \caption{Sketch of a segment of a reflection-symmetric sinusoidally varying channel that is confining the motion of overdamped, point-like
Brownian particles. The modulated sidewalls $\pm \omega (x)$ are $L$-periodic, $\Delta H$ represents the channel height, and
the minimal and maximal channel widths are $\Delta \omega$ and $\Delta \Omega$, respectively.
The size of an unit cell is indicated with the dashed lines. Superimposed is an exemplary force
field $\V{F}(\V{q})$ which contains vortices and stagnation points (solid circles) at the central $z$-layer, i.e., $z=\Delta H/2$.}
  \label{fig:Fig1}
\end{figure}

Besides the direct forcing of the particle dynamics, the use of hydrodynamical flows presents an additional ``degree of freedom'' to control particle transport \cite{Schindler2007,Bechinger2011,Scholz2012,Scholz2012b,Ai2013,Ai2014,Loewen2014} and optimize separation efficiency, speed and purity \cite{Martens2013b,Eijkel2005}. As pointed out in recent own work \cite{Martens2013}, upon combining a constant force causing the particle to move along the channel and a pressure-driven flow that drags the particle in the opposite direction results in the phenomenon of {\itshape hydrodynamically enforced entropic trapping} (HEET); the latter implies that for certain values of the constant force and the pressure drop, the mean particle velocity vanishes and, in addition, the particles' diffusivity is reduced significantly. In Ref.~\cite{Martens2013b}, the mechanisms of \textit{entropic rectification} is revealed in channels with a broken spatial reflection symmetry in presence of HEET. We demonstrated
that efficient rectification with a drastically reduced diffusivity can be achieved due to the combined action of the forcing and the pressure-driven flow field. In both those works, we applied the general three-dimensional theory  to two-dimensional channels (i.e. $\Delta H \gg \Delta \Omega$ in \figref{fig:Fig1}) and thus neglected the impact of the channel height on the hydrodynamical flows. In this work, however, we study the impact of the channel height which is limited in microfluidic devices on the particle transport.

After presenting the model in \secref{sec:model}, we apply the generalized Fick-Jacobs theory to a three-dimensional channel geometry with sinusoidally modulated walls in \secref{sec:fickjacobs}.
We derive analytic estimates for the flow velocity in corrugated channels in \secref{subsect:pois}. \secref{sec:HEET}
is devoted to the combined action of constant forcing and the hydrodynamical flow field on the transport quantities like the mean particle velocity and the effective diffusion coefficient. In particular, we discuss the impact of the channel height on the HEET effect. In \secref{sec:conclusion} we summarize our main findings.


\section{Brownian particles in microfluidic channel systems}

\subsection{Statement of the problem} \label{sec:model}

We start by considering spherical Brownian particles of radius $R$ suspended in an aqueous solvent of density $\rho$ and dynamic viscosity $\eta$. The latter fills a planar, three-dimensional, $L$-periodic channel with confining periodic walls at $y=\omega_{\pm}(x)$ and plane walls placed at $z=0$ and $z=\Delta H$, see in \figref{fig:Fig1}. The maximum and minimum widths in $y$-direction are $\Delta\Omega$ and $\Delta\omega$, respectively. Assuming that (i) the particle suspension is dilute and (ii) particles are small, $R \ll \Delta\omega$, with density comparable with $\rho$, inertial effects, hydrodynamic particle-particle and particle-wall interactions, and the effects that can be initiated by rotation of particles can be neglected \cite{Happel1965,Maxey1983}. If so, the particles evolve in a laminar flow and the dynamics is well described by the overdamped Langevin equation. By measuring the lengths, $\V{q} \to \V{q}\, L$, energies, $\Phi \to \Phi\,k_BT$ and $\grb{\Psi} \to \grb{\Psi}\,k_BT$, forces,
$\V{F} \to \V{F}\, k_BT/L$, and time, $t \to \tau\, t$, in the scales of the period $L$, the thermal energy $k_BT$, and the relaxation time $\tau=\, 6\pi \eta R L^2/(k_B T)$, respectively, we arrive at the dimensionless Langevin equation describing the overdamped motion of Brownian particles \cite{Martens2013}, i.e.,
\begin{align}
\pt{\V{q}}{t} =\V{F}(\V{q}) + \sqrt{2}\,\grb{\xi}(t),
\label{eq:eom}
\end{align}
where ${\V{q}=(x,y,z)^T}$ is the particle position and the Gaussian random force $\grb{\xi}=(\xi_x,\xi_y,\xi_z)^T$ obeying
$\av{\xi_i(t)} = 0$, $\av{\xi_i(t) \xi_j(s)}=\delta_{ij}\delta(t-s)$; $i,j \in \left\{x,y, \mbox{or}\, z\right\}$.

Generally, any force field $\V{F}$ exerted on particles can be decomposed into a curl-free part (described by a scalar potential $\Phi$) and a divergence-free part (described by a vector potential $\grb{\Psi}$), which constitute the two components of the Helmholtz's decomposition theorem,
\begin{align}
 \V{F}(\V{q})=-\grad \Phi(\V{q})+\grad\times\grb{\Psi}(\V{q}). \label{eq:forcefield}
\end{align}
We note that although the force $\V{F}(\V{q})$ can be very general, we require that either ${\V{n}\cdot(\grad\times\grb{\Psi})=0}$ or
${\V{n}\times \grad \Phi=0}$ is fulfilled at the boundaries. Hereby, the outward-pointing normal
vector is $\V{n}=(\mp \omega_\pm'(x),\pm 1,0)^T$ at the side-walls and $\V{n}=(0,0,\pm 1)^T$ at the top and bottom boundary, respectively. This ensures that the harmonic component, which can generally be non-zero for bounded domains (the Hodge or Helmholtz-Hodge decomposition) vanishes, see discussion in Sec. 3 of Ref.~\cite{Bhatia2013}. For the fluidic system considered in this work, the fact that the flow vanishes at the walls guarantees that the first of the above boundary requirements is always met.


Hereafter, we focus on the interplay of curl-free and divergence-free forces on the particle transport. On the one hand, we consider an external constant bias in $x$-direction with magnitude $f$, leading to $\Phi(\V{q})=-f\,x$. On the other hand, we account for the difference between the particle velocity
$\dot{\V{q}}$ and the local instantaneous velocity of the solvent $\V{u}(\V{q})$ based on the Stokes law, which gives us
$\V{u}(\V{q}) = \grad\times\grb{\Psi}(\V{q})$ with $\grb{\Psi}(\V{q}) = \Psi(\V{q}) \V{e}_z$ \cite{Martens2013,Martens2013b}.
Here, $\Psi(\V{q})$ is the hydrodynamic stream function. As a result, \eq{eq:eom} turns into the Langevin equation
\begin{align}
  \pt{\V{q}}{t}=\,f\V{e}_x+\V{u}(\V{q})+\sqrt{2}\,\grb {\xi}(t),
\label{eq:eom_flow}
\end{align}
to be supplemented by {\it no-flux} boundary conditions for the particles at the walls. We stress that our consideration completely neglects particle feedback effects, meaning that the fluid flow is not affected by the particles and the flow field can be simply superimposed \cite{Straube2011}.


\subsection{\label{sec:fickjacobs} Generalized Fick-Jacobs approach}

We next present our \textit{generalized Fick-Jacobs approach} \cite{Martens2013} which extends the standard Fick-Jacobs theory
towards the most general force fields $\V{F}(\V{q})$ as detailed with \eq{eq:forcefield}. The essential task entails computing the probability density
function (PDF) $P\bracket{\V{q},t}$ of finding the particle at the local position $\V{q}$ at time $t$. The evolution of $P\bracket{\V{q},t}$ is determined by the
Smoluchowski equation \cite{Burada2009_CPC}
\begin{align}
0=&\,\partial_t P\bracket{\V{q},t}+\grad \cdot \V{J}\bracket{\V{q},t}, \label{eq:smol}
\intertext{with probability flux}
\V{J}\bracket{\V{q},t}=&\,\V{F}(\V{q})\,P\bracket{\V{q},t}-\grad\,P\bracket{\V{q},t}.
\end{align}
Caused by the impenetrability of the channel walls, the probability flux
obeys the no-flux boundary condition at the walls,
viz., $\V{J}\bracket{\V{q},t}\cdot\V{n}=0$. 
The prime denotes the derivative with respect to $x$. Moreover, the PDF obeys the initial condition $P\bracket{\V{q},0}=P_\mathrm{init}\bracket{\V{q}}$ at time $t=0$.

In the spirit of the Fick-Jacobs approach \cite{Jacobs,Zwanzig1992,Burada2009_CPC}, i.e.,
assuming fast equilibration in the transverse channel direction, we perform an asymptotic perturbation analysis in the geometric parameter
$\e = \bracket{\Delta\Omega-\Delta\omega}/L \ll 1$ \cite{Martens2011,Martens2011b,Laachi2007}.
Upon re-scaling the transverse coordinate $y\to \e\,y$, we expand the joint PDF
in a series in even powers of $\e$, viz., $P(\V{q},t)=P_0(\V{q},t)+\e^2\,P_1(\V{q},t)+\mathcal{O}(\e^4)$, and similarly for $\Phi(\V{q})$ and $\grb{\Psi}(\V{q})$, respectively. Substituting these expansions into \eq{eq:smol} and taking account of the the boundary conditions, where we explicitly claim that $\bracket{\grad\times\grb{\Psi}_0(\V{q})}_z$ vanishes at the upper $z=\Delta H$ and lower confining boundary $z=0$, the kinetic equation for the time-dependent marginal PDF $P_0(x,t)=\int_{\omega_-(x)}^{\omega_+(x)}\mathrm{d}y\int_{0}^{\Delta H}\mathrm{d}z P_0(\V{q},t)$ is the \textit{generalized Fick-Jacobs equation}, which reads
\begin{align} \label{eq:genFJ}
  \frac{\partial}{\partial t} P_0(x,t) = \frac{\partial}{\partial x}
\left[ \left( \frac{\mathrm{d} \mathcal{F}(x)}{\mathrm{d} x} \right)
P_0\right]+\frac{\partial^2}{\partial x^2} P_0.
\end{align}
Here $\mathcal{F}(x)$ is the \textit{generalized potential of mean force} given by
\begin{align}
 \label{eq:meanforce}
 \mathcal{F}(x)\!=\!-\ln\left[\intl{h_-(x)}{h_+(x)}{y}\!\!\intl{0}{\Delta H}{z}\!
e^{-\Phi_0}\!\right] -\intl{0}{x}{x'}\!\!\intl{h_-(x')}{h_+(x')}{y}\!\!\intl{0}{\Delta H}{z}\!
\bracket{\grad \times \grb{\Psi_0}}_x\!P_\mathrm{eq}(y,z|x'),
\end{align}
where $P_\mathrm{eq}(y,z|x)=\,e^{-\Phi_0(\V{q})}\Big/\int_{h_-(x)}^{h_+(x)}\mathrm{d}y\int_{0}
^{H} \mathrm{d}z\, e^{-\Phi_0(\V{q})}$  is the equilibrium PDF of $y$ and $z$,
conditioned on $x$. The derivation of Eqs.~\eqref{eq:genFJ} and \eqref{eq:meanforce} is presented in detail in our
Appendix~\ref{app:genFJ}.

We note that $\mathcal{F}(x)$ comprises the usual \textit{entropic} contribution \cite{Reguera2001,Burada2008}, i.e. the first, logarithmic term,  which is caused by the non-holonomic constraint stemming from the boundaries \cite{Sokolov2010} and the newly energetic contribution \cite{Martens2013}, i.e. the part stemming from  ${\grb{\Psi}_0}$, which is associated with the conditional average of the $x$-component of divergence-free forces exerted on the particle weighted by its equilibrium conditional PDF $P_\mathrm{eq}(y,z|x)$. In the absence of $\grb{\Psi}$, Eqs.~\eqref{eq:genFJ} and \eqref{eq:meanforce} reduce to the commonly known result of the \textit{Fick-Jacobs} approximation \cite{Zwanzig1992}.

In the long time limit, $\lim_{t\to \infty}
P\bracket{\V{q},t}=P\bracket{\V{q}}$, the PDF has to satisfy the normalization condition $\int_{\mathrm{unit-cell}}^{}
P(\V{q})\, \mathrm{d}^3\V{q} =1$, and be periodic, $P(x+m,y,z)=P(x,y,z)\,,\forall m \in \Z$. Referring to \eq{eq:genFJ}, the stationary marginal PDF reads
\begin{align}
\label{eq:pdf}
P_0(x)=\mathcal{I}^{-1}I(x),
\end{align}
in leading order, with $I(x) = e^{-{\cal F}(x)} \int_{x}^{x+1}\mathrm{d}x' e^{{\cal F}(x')}$ and $\mathcal{I} = \int_{0}^{1} \mathrm{d}x\,I(x)$.

Note that a closed-form expression for $P_0(x)$ exists only if the scalar potential $\Phi_0$ is either independent of the $x$-coordinate or scales linearly
with $x$, and if and only if the longitudinal coordinate $x$ is not multiplicative connected to the transverse coordinates. Further, \eq{eq:pdf} is only valid if the $x$-component of $\bracket{\nabla \times \grb{\Psi}_0}_x$ is periodic in $x$ with unit period an the
generalized potential of mean force fulfills the condition $\Delta \mathcal{F}=\mathcal{F}(x+1)-\mathcal{F}(x)\neq 0$. For $\Delta\mathcal{F}=0$, the stationary
joint PDF is constant, ${P_0(\V{q})=const}$, and the marginal PDF scales with the local channel cross-section ${P_0(x)\propto Q(x)}$, wherein $Q(x)=\,\Delta H\bracket{\omega_+(x)-\omega_-(x)}$.

We evaluate the stationary average particle current
by use of well-known analytic expressions \cite{Burada2009_CPC}, to yield
\begin{align}
 \av{\dot{x}}=&\,\mathcal{I}^{-1}\,\bracket{1-e^{\Delta\mathcal{F}}}. \label{eq:currFJ}
\end{align}
The effective diffusion coefficient
$D_\mathrm{eff} = \lim_{t\to\infty} (\av{x^2(t)}-\av{x(t)}^2)/(2t)$ (in units of the bulk diffusivity, $D_0=k_BT/(6 \pi\eta R)$) is
calculated via the first two moments of the  first passage time distribution, see Eq. (17) in
Ref.~\cite{Burada2009_CPC}, leading to
\begin{align}
D_\mathrm{eff} &= \mathcal{I}^{-3}\, \int_{0}^{1} \mathrm{d}x \, \int_{x-1}^{x} \mathrm{d}x' \, e^{{\cal F}(x) - {\cal
F}(x')}\, I^2(x). \label{eq:effectivediffusion}
\end{align}

\section{Poiseuille flow in a weakly shape-perturbed three-dimensional channel}

\subsection{\label{subsect:pois} General solution for arbitrary profile functions}

We are interested in the stationary fluid flow through a three-dimensional, planar channel geometry with periodically varying cross-section as, e.g., in Fig.~\ref{fig:Fig1}. A slow pressure-driven steady flow of an incompressible solvent is determined by the dimensionless Stokes equations \cite{Happel1965,Bruus2008}
\begin{align}
\grad \mathcal{P}(\V{q})=\grad^2 \V{u}(\V{q}), \qquad \grad
\cdot \V{u}\bracket{\V{q}}=0, \label{eq:SE}
\end{align}
being valid for small Reynolds number ${\rm Re}\ll 1$. Here, the flow velocity $\V{u}=\bracket{u^x,u^y,u^z}^T$ and the pressure $\mathcal{P}(\V{q})$ are measured in the units of $L/\tau$ and $\eta/\tau$, respectively. We require that $\V{u}$ is periodic, $\V{u}(x,y,z)=\V{u}(x+1,y,z)$, and obeys the no-slip boundary conditions,
$\V{u}(\V{q})=0$, $\forall\, \V{q} \in \mbox{channel wall}$. The pressure is supposed to satisfy the jump condition $\mathcal{P}(x+1,y)=\mathcal{P}(x,y)+\Delta \mathcal{P}$ where $\Delta \mathcal{P}$ is the pressure drop along one unit cell.

By looking for a plane parallel flow in the form $\V{u}(\V{q})=(u^x(\V{q}),u^y(\V{q}),0)^T$, we find that the local pressure is independent of $z$, $\mathcal{P}=\mathcal{P}(x,y)$. For the flow confined between the solid plane walls at $z=0, \Delta H$, satisfying the no-slip requirement at these boundaries and being consistent with \eq{eq:SE}, the flow field can be represented as series \cite{Bruus2008}
\begin{align}
(u^x(\V{q}),u^y(\V{q}))^T=&\,\sum_{m=0}^\infty (f_{m}(x,y),g_{m}(x,y))^T\,\sin\left(\frac{c_m z}{\Delta H}\right), \label{eq:app_series}
\end{align}
yielding the relation
\begin{align}
\frac{4 (\partial_x \mathcal{P}, \partial_y \mathcal{P})^T}{c_m}=&\left(\bigtriangleup_\mathrm{2D} -\frac{c_m^2}{\Delta H^2}\right)(f_{m},g_{m})^T, \label{eq:series_z}
\end{align}
where $c_m=(2m+1)\pi$ and $\bigtriangleup_\mathrm{2D}=\partial_x^2+\partial_y^2$; for the $z$-independent pressure gradient terms we have used the representation of a constant $1=\sum_{m=0}^{\infty}4\, c_m^{-1}\sin(c_m z/\Delta H)$.

In the full analogy to the derivation of the generalized Fick-Jacobs equation, all transverse
quantities are measured in units of expansion parameter $\e$, i.e., $y\to\e\,y$, ${\omega_\pm(x)\to \e\,h_\pm(x)}$, and $g_m\to\e\,g_m$.
The existence of a Poiseuille flow solution in a channel with non-deformed wall profiles at $y=\omega_{\pm}=const$, additionally requires a typical rescaling of pressure, $\mathcal{P}\to\e^{-2}\,\mathcal{P}$. Afterwards, we perform the long-wave expansion
\begin{align}
 (f_{m},g_{m},\mathcal{P})^T = \sum_{n=0}^\infty \e^n (f_{m}^{(n)},g_{m}^{(n)},\mathcal{P}_{n})^T. \label{eq:series}
\end{align}
After the substitution into \eq{eq:series_z}, we obtain
\begin{subequations}
 \begin{align}
  \frac{4}{c_m}\partial_x \mathcal{P}_0 & =\partial_y^2
f_m^{(0)}-\kappa^2 c_m^2  f_m^{(0)},\label{eq:leadorder1} \\
  \frac{4}{c_m} \partial_y \mathcal{P}_0 & =0\,, \quad \mbox{with}\quad \kappa=\frac{\e}{\Delta H},
\label{eq:leadorder2}
 \end{align}\label{eq:leadorder}%
\end{subequations}
in the leading order in $\e$.
Note that the parameter $\kappa$ is a ratio of two parameters; both can be arbitrarily small. While for channels with wide gaps between the walls at $z=0$ and $z=\Delta H$, $\kappa \ll 1$ ($\Delta H \gg \e$) and thus the third term in \eq{eq:leadorder1} becomes negligible, this term dominates for small gaps $\kappa \gg 1$ ($\Delta H \ll \e$); the latter is referred to as the Hele-Shaw flow limit.

It follows from \eq{eq:leadorder2} that the pressure $\mathcal{P}_0$ depends only
on the $x$-coordinate and we can integrate \eq{eq:leadorder1}, whose general solution reads
\begin{align}
 f_m^{(0)}=\,A_m\,\sinh\bracket{c_{m} \kappa y}+B_m\,\cosh\bracket{c_{m}
\kappa y}-\frac{4\partial_x \mathcal{P}_0(x)}{c_{m}^3 \kappa^2}.
\end{align}
Satisfying the no-slip boundary conditions $f_m(x,y=h_\pm(x))=0$, we find the constants $A_m$ and $B_m$ to arrive at
\begin{align} \label{eq:ux_rect}
\begin{split}
  u_0^x(\V{q})=\,-\frac{4 }{\kappa^2}\partial_x \mathcal{P}_0(x)\,\sum_{m=0}^\infty
&\left\{1-\frac{\sinh\left[c_m \kappa \left(h_+(x)-y\right)\right]}{\sinh\left[c_m \kappa\mathcal{H}(x)\right]}
-\right. \\ &\left. - \frac{\sinh\left[c_m \kappa (y-h_-(x))\right]}{\sinh\left[c_m \kappa \mathcal{H}(x)\right]}
\right\} c_m^{-3} \sin \left(\frac{c_{m}z}{\Delta H}\right),
\end{split}
\end{align}
where $\mathcal{H}(x)=h_+(x)-h_-(x)$ is the re-scaled local width. From the integrated continuity condition follows that the transverse velocity component $u_0^y(\V{q})$ is determined by $g_m^{(0)}(x,y)=-\partial_x \int_{h_-(x)}^y f_m^{(0)}(x,y')\,{\rm d}y'$. Satisfying the no-slip conditions for $u_0^y(\V{q})$ at $y=h_{\pm}(x)$ allows us to obtain the local pressure $\mathcal{P}_0(x)$,
\begin{align}
 \mathcal{P}_0(x)=\,\mathcal{P}^0+\frac{\Delta \mathcal{P}}{\av{\chi^{-1}(x)}_x}\, \intl{0}{x}{x'}\,
\chi^{-1}(x'),\label{eq:pres_rect}
\end{align}
where
\begin{align}
\chi(x)=\frac{1}{\kappa^3}\left[\frac{7 \zeta(3)}{8\pi^3}\kappa\mathcal{H}(x)+2\sum_{m=0}^\infty \frac{1-\cosh\bracket{c_m \kappa \mathcal{H}(x)}}{c_m^4 \sinh(c_m \kappa \mathcal{H}(x))}\right]. \label{eq:preschi}
\end{align}
Here, $\Delta \mathcal{P}$ is the drop of pressure along one unit-cell, $\mathcal{P}^0$
corresponds to a constant offset which can be set to zero and $\zeta(x)$ is the Riemann zeta function.

Note that for reflection symmetric cross-sections, $h_\pm(x)=\pm h(x)$, \eq{eq:ux_rect} resembles Eq.~(9) in \cite{Lauga2004} where the leading order solution was obtained via lubrication theory, using the channel height as an expansion parameter, $\Delta H\ll 1$.

\subsection{\label{subsect:pois-sym-corr} Poiseuille flow in a reflection-symmetric geometry}

\begin{figure}
\centering
\includegraphics[width=0.49\linewidth]{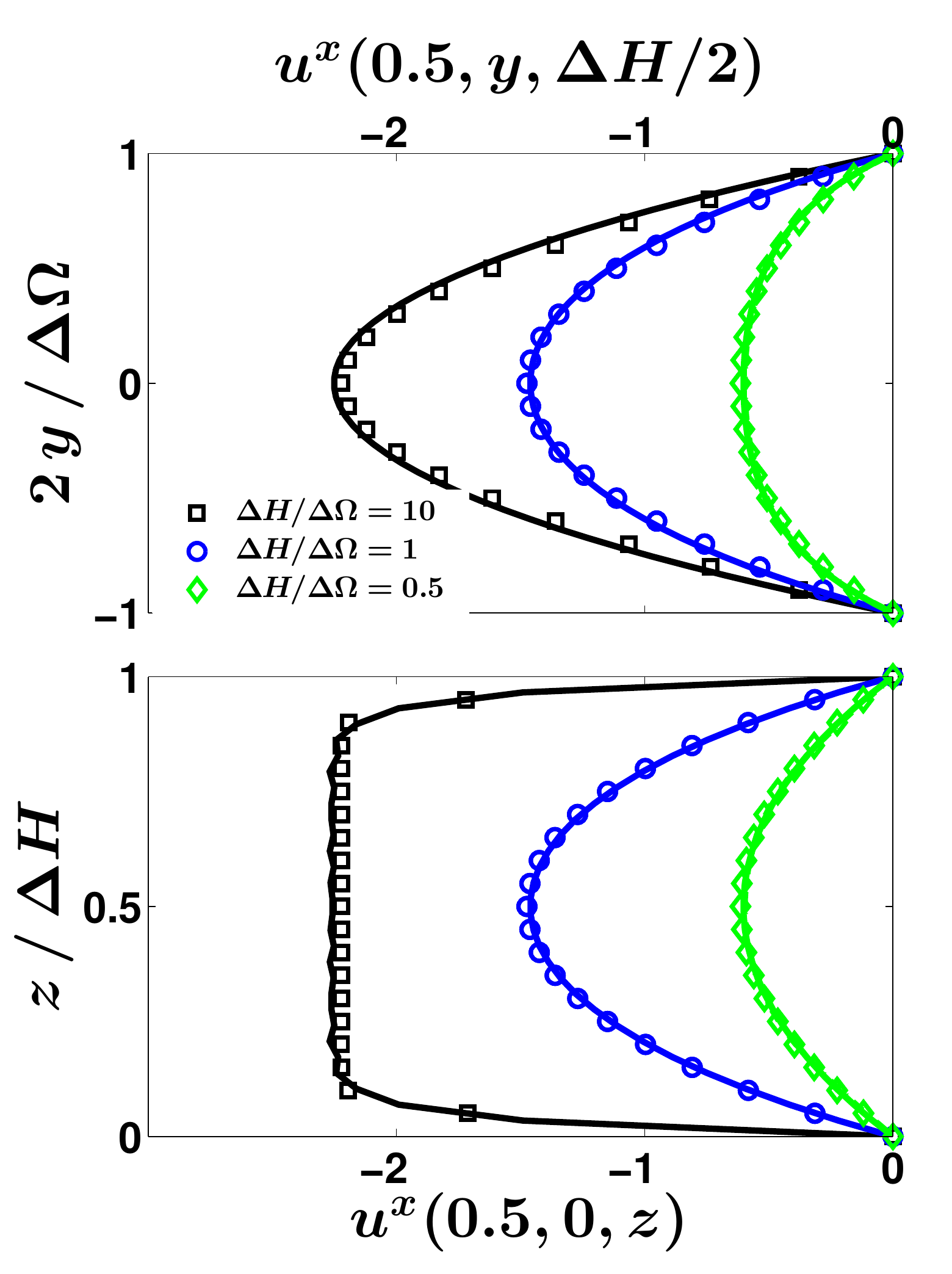}
\includegraphics[width=0.49\linewidth]{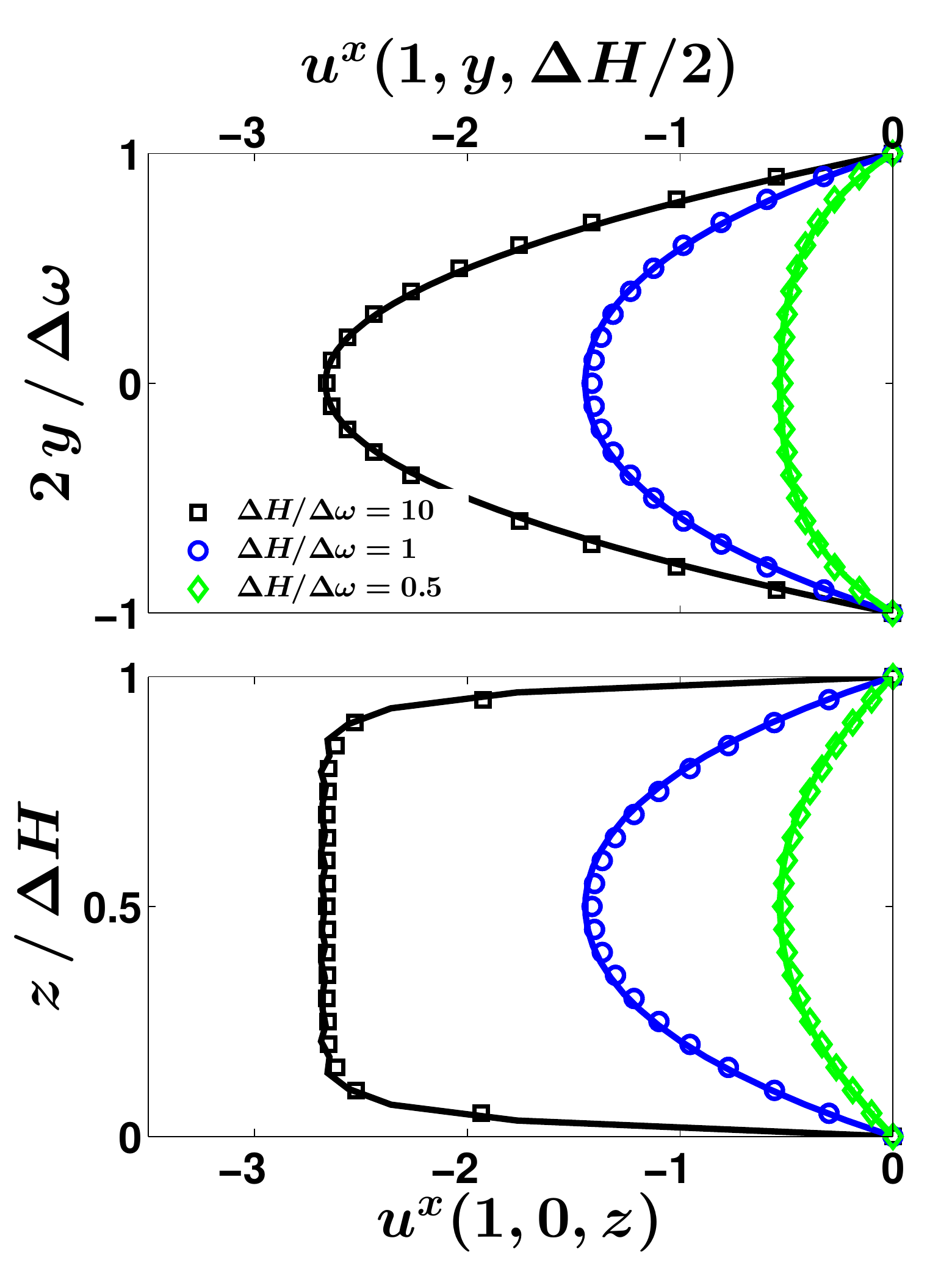}
  \caption{Profiles of the longitudinal flow component $u^x$ in a $3$D channel with sinusoidally modulated boundary,
$\omega_\pm(x)$, \eq{eq:conf}, and constant height $\Delta H$, see \figref{fig:Fig1}. The profiles in the $y-z$ plane were numerically evaluated by FEM (markers) for
different ratios of channel height to local width: $\Delta H/\Delta \Omega$ at $x_\mathrm{pos}=0.5$ (left column) and
$\Delta H/\Delta \omega$ at $x_\mathrm{pos}=1$ (right column). The solid lines represent the analytic estimate,
\eq{eq:ux_rect}, by calculating the sum of the first $10$ terms. The remaining parameter values are $\Delta\Omega=0.5$, $\Delta\omega=0.4$, and $\Delta \mathcal{P}=100$.}
\label{fig:Fig2}
\end{figure}

In \figref{fig:Fig2} we depict the numerical solution of the Stokes equation in a three-dimensional channel, \eq{eq:SE}, with reflection symmetric
sinusoidally-shaped sidewalls \cite{Martens2011,Martens2011b}, cf.~\figref{fig:Fig1}, using finite element method (FEM)
\begin{align}
 \omega_\pm\bracket{x}=&\,\pm\left[\frac{\Delta\Omega+\Delta\omega}{4}-\frac{
\Delta\Omega-\Delta\omega}{4}\cos\bracket{2\pi x}\right]. \label{eq:conf}
\end{align}
It is shown that the flow profile in both $y$- and $z$- directions is symmetric with respect to the the center of the channel. The maximal flow velocity increases with growing channel height whereby the pressure-driven flow attains its largest value at the bottleneck due to the continuity condition $\nabla\cdot \V{u}=0$. We note that the profiles $u^x(\V{q})$ in both $y$- and $z$-directions crucially depend on the channel height $\Delta H$. While $u^x(\V{q})$ remains highly parabolic in $z$ for very thin channels $\Delta H\ll \Delta \Omega$, the profile becomes flat in $z$ except near the walls in the limit of infinite channel $\Delta H \gg \Delta\Omega$. The behavior of $u^x$ in $y$-direction has an opposite tendency. The profile is parabolic for all $\Delta H$ except for the case of very thin channels, $\Delta H \ll \Delta \Omega$, when the profile of $u^x$ as a function of $y$ becomes flatter (not depicted in \figref{fig:Fig2}).

We find that our analytic expression for the longitudinal flow component, \eq{eq:ux_rect}, agrees very well with the
numerics for weakly modulated channels, cf. \figref{fig:Fig2} ($\e=0.1$). With growing channel corrugation, i.e., $\e \to \Delta\Omega$, respectively, $\Delta\omega \to 0$, the leading order theoretical prediction tends to overestimate the numerical results, especially at the bottleneck. Nevertheless, the agreement remains notably well at the widest part of the channel (not explicitly shown).

\subsection{Thin channel -- Hele-Shaw limit}

\noindent As seen from the numerical results, when the gap between the two plane walls at $z=0$ and $z=\Delta H$ is small, the velocity profile in the $z$-direction is parabolic, see \figref{fig:Fig2}. This is consistent with the known asymptotic relation between the flow velocity and the pressure gradient
\begin{align}
 \V{u}=-\frac{1}{2}\nabla \mathcal{P}(x,y)\,z\bracket{\Delta H-z}, \label{eq:HS}
\end{align}
which formally holds in the Hele-Shaw limit $\Delta H \to 0$. The latter and the uniformity of the pressure in the $z$-direction allows us to integrate $\V{u}$ with respect to $z$ and thus to consider an effective velocity field in only the two dimensions $x$ and $y$. Substituting \eq{eq:HS} into the continuity equation, $\nabla\cdot \V{u}=0$,  and integrating the latter over $z$, we obtain the governing equation of Hele-Shaw flows
\begin{align}
 0=\,\bigtriangleup_\mathrm{2D} \mathcal{P}=\,\left[\partial_x^2+\partial_y^2\right] \mathcal{P}(x,y).
\end{align}
This equation is supplemented by the no-slip boundary conditions on the side walls of the channel, $\nabla \mathcal{P} \cdot \V{n}=0$, where $\V{n}=(\mp \omega_\pm'(x),\pm 1,0)^T$ is a unit vector perpendicular to the side wall and the discussed jump condition $\mathcal{P}(x+1,y)=\mathcal{P}(x,y)+\Delta \mathcal{P}$.

As earlier, if we rescale the transverse coordinate $y\to \e\,y$ and expand the local pressure in a series in even powers of $\e$,  $\mathcal{P}=\mathcal{P}_0+\e^2\mathcal{P}_1+\mathcal{O}(\e^4)$, in the leading order we find
\begin{subequations}
\begin{align}
 \mathcal{P}_0(x,y)=\,\frac{\Delta\mathcal{P}}{\av{W(x)^{-1}}_x}\,\intl{0}{x}{x'} W(x')^{-1},
 \intertext{and the longitudinal flow component becomes}
 u_0^x=\,-\frac{\Delta\mathcal{P}}{2\,\av{W(x)^{-1}}_x}\,\frac{z\bracket{\Delta H-z}}{W(x)}. \label{eq:ux0_HS}
\end{align}
\end{subequations}
Here, the results are written in terms of $W(x)=\omega_+(x)-\omega_-(x)$.

\subsection{Limit of a high channel}

\noindent For microfluidics devices the aspect ratio of height to width can often be so large that $\Delta H \gg \Delta\Omega$. In this case, the velocity profile for $u^x$ in $z$ is flat, cf. square markers in \figref{fig:Fig2}. Taking the limit $\kappa \to 0$ in \eqs{eq:ux_rect}-\eqref{eq:preschi}, we obtain the quasi-two dimensional results \cite{Martens2013}
\begin{subequations}
\begin{align}
u^x_0(x,y)\simeq&\,-\frac{\partial_x\mathcal{P}_0(x)}{2}\bracket{\omega_+(x)-y}\,\bracket{y-\omega_-(x)}, \label{eq:ux0_2D}\\
 \mathcal{P}_0(x,y)\simeq&\,\frac{\Delta \mathcal{P}}{\av{W(x)^{-3}}_x}\,\intl{0}{x}{x'} W(x')^{-3}.
\end{align}
\end{subequations}
Here, we have used the definition of $\kappa$ and the relation $\varepsilon\mathcal{H}(x)=W(x)$.

Interestingly, irrespective of the channel height the local pressure sharply increases, respectively, decreases at the channel's bottlenecks instead of the linear growth known for straight channels. Consequently, $u_0^x$ attains its maximum values at the bottlenecks where the gradient of the local pressure is maximal. Thereby, the longitudinal flow component $u_0^x$ is inversely proportional to the third power of the local channel width $W(x)$ for infinite high channels and only proportional to $1/W(x)$ for very thin channels. From this reasoning we can conclude that the flow velocity in longitudinal direction is limited from above by the values
\begin{subequations}
 \begin{align}
 \mathrm{max}(|u_0^x|)\leq&\, |\Delta\mathcal{P}|\, \frac{(\Delta H)^2 \Delta\Omega}{8 \Delta\omega},\quad \mbox{for}\quad \Delta H\to 0, \label{eq:umax_HS}\\
  \mathrm{max}(|u_0^x|)\leq&\, |\Delta\mathcal{P}|\,\frac{(\Delta\Omega)^3}{8\,\Delta\omega},\quad \mbox{for}\quad \Delta H\to \infty.
 \end{align}
\end{subequations}

\section{\label{sec:HEET} Hydrodynamically enforced entropic trapping}

Next, we investigate the dependence of the transport quantities, such as the
average particle velocity $\av{\dot{x}}$, \eq{eq:currFJ}, and the effective diffusion coefficient
$D_\mathrm{eff}$, \eq{eq:effectivediffusion}, on the force magnitude $f$ and the pressure drop $\Delta \mathcal{P}$
which control the curl-free and the divergence-free contributions in \eq{eq:forcefield}, respectively.

\begin{figure}
 \includegraphics[width=0.5\linewidth]{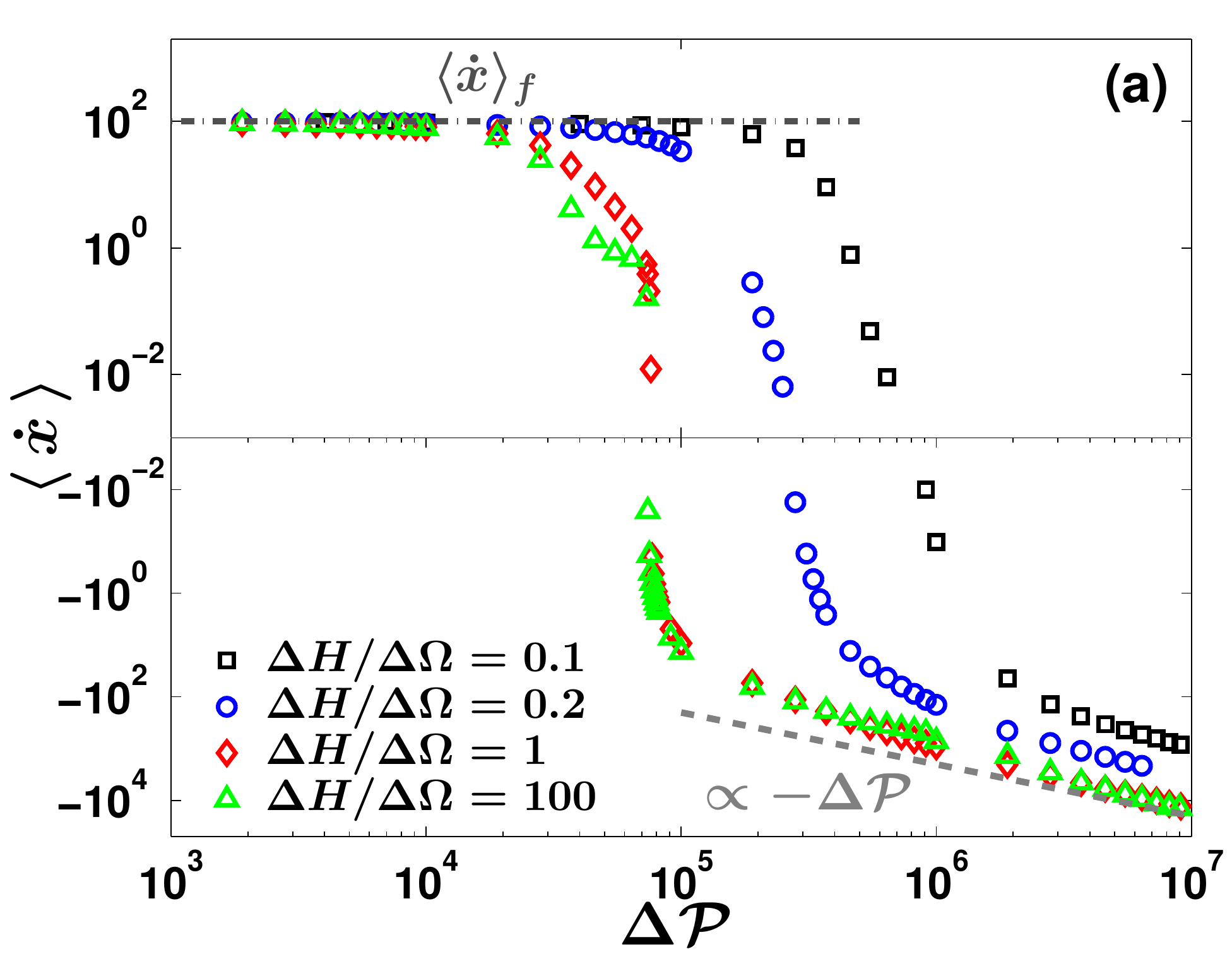}
 \includegraphics[width=0.5\linewidth]{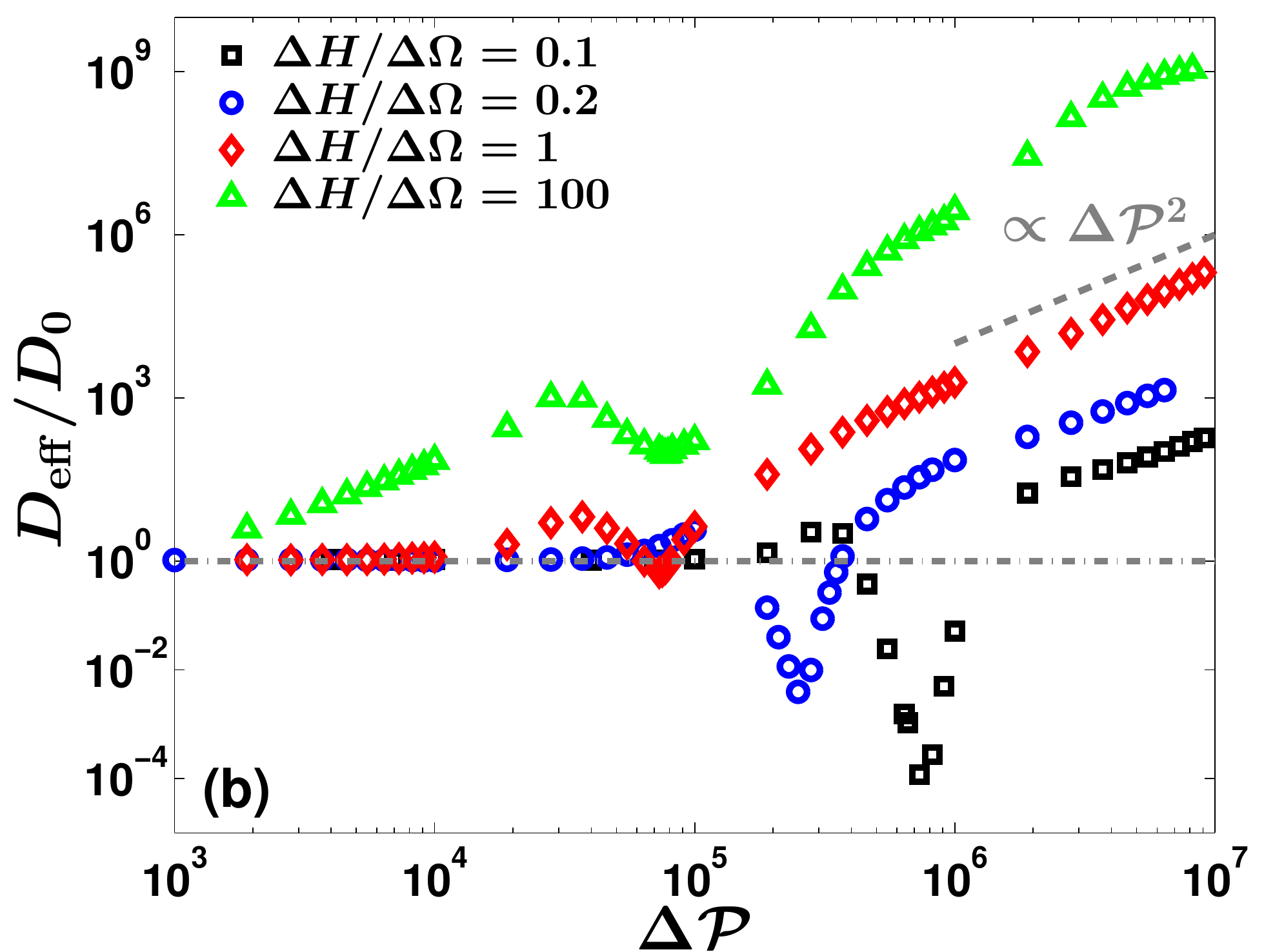}
  \caption{
 The scaled average particle velocity $\av{\dot{x}}$ (a) and the effective diffusion coefficient $D_\mathrm{eff}$ (b) as functions
 of the pressure drop $\Delta\mathcal{P}$ for different channel heights $\Delta H$.
 In panel (a): The dash-dotted line  corresponds to $\Delta \mathcal{P}\simeq 0$, \eq{eq:vel_entropic}, and the dashed line represent the asymptotic limit $\av{\dot{x}}\propto - \Delta \mathcal{P}$ for $\Delta \mathcal{P} \gg f$. In panel (b): The horizontal dash-dotted line corresponds to the value of the bulk diffusivity. The markers are for Brownian dynamics simulations of \eq{eq:eom_flow} with the flow field obtained via numerically solving \eq{eq:SE}. For all simulation results the external force magnitude is kept fixed $f=100$ and the maximum and minimum channel width are set to $\Delta\Omega=0.5$ and $\Delta\omega=0.1$, respectively.} \label{fig:Fig3}
\end{figure}

In \figref{fig:Fig3}(a), we depict the mean particle velocity $\av{\dot{x}}$ as a function of the applied pressure drop
$\Delta {\mathcal P}$ for different channel heights $\Delta H$. The strength of the external bias is kept fixed, viz., $f=100$. The numerical results are obtained by Brownian dynamics simulations of \eq{eq:eom_flow} for $3\cdot 10^4$ particles. While $\av{\dot{x}}$ exhibits a linear dependence on $\Delta \mathcal{P}$ and, consequently, it is point symmetric with respect to $\Delta \mathcal{P}$ only in the case of purely flow induced transport ($f=0$), its behavior changes drastically for $f \neq 0$. For small to intermediate values of $|\Delta {\mathcal P}|$, the presence of the flow is insignificant and thus the mean particle velocity remains almost constant. In this interval the external bias dominates and hence $\mathcal{F}(x)$ is approximately given by $\mathcal{F}(x)\simeq\,-f\,x-\ln[2\omega(x)]$. According to \eq{eq:currFJ}, the mean particle current reads
\begin{align}
\av{\dot{x}}_f\simeq \frac{f^3+4\pi^2\,f}{f^2+2 \pi^2 (\sqrt{\delta}+1/\sqrt{\delta})},\quad \mbox{for}\;\;\Delta \mathcal{P}=0\,, \label{eq:vel_entropic}
\end{align}
where $\delta= \Delta\omega/\Delta\Omega$. For $\Delta {\mathcal P}>0$, the solvent flow drags the particles into the direction opposite to the external force ($u^x<0$ and $f>0$). At a critical pressure drop $\Delta \mathcal{P}_\mathrm{cr}$, a sharp transition of $\av{\dot{x}}$ from positive to negative values occurs, cf. Refs.~\cite{Martens2013,Martens2013b}. Upon further increasing $\Delta \mathcal{P}$, the flow-induced Stokes' drag force starts to dominate over the static bias $f$ and thus $\av{\dot{x}} \propto -\Delta \mathcal{P}$, see the dashed line in \figref{fig:Fig3}(a).

Although strong nonvanishing local forces $f\V{e}_x+\V{u}(\V{q})$ are acting and the particles experience continuous thermal fluctuations, there exists a critical ratio $\bracket{f/\Delta \mathcal{P}}_\mathrm{cr}$ such that the mean particle velocity vanishes identically $\av{\dot{x}}=0$. We observe that the value of $\Delta \mathcal{P}_\mathrm{cr}$ crucial depends on the channel height $\Delta H$ for a given sidewall profile, cf. \figref{fig:Fig3}(a). According to \eq{eq:umax_HS}, the maximum value $\mathrm{max}(|u_0^x|)$ scales with $\Delta \mathcal{P}\,\Delta H^2$ for thin microfluidic channels. Consequently, the thinner the channel the larger is the needed pressure difference in order to reverse the particle's transport direction for a given value of $f$. Using our approximated results for longitudinal flow components \eq{eq:ux0_HS} for \textit{Hele-Shaw} like channels and \eq{eq:ux0_2D} for quasi $2$D channels, we can estimate the lower and upper limit for the critical ratio  $\bracket{f/\Delta \mathcal{
P}}_\mathrm{cr}$. As follows from \eq{eq:currFJ}, $\av{\dot{x}}=0$ if and only if $\mathcal{F}(x+1)-\mathcal{F}(x)=\Delta \mathcal{F}=0$, yielding
\begin{align}
 \bracket{\frac{f}{\Delta \mathcal{P}}}_\mathrm{cr}=&\,\frac{\Delta H^2}{12},\quad \mbox{for}\,\Delta H\to 0, \label{eq:f_dp_crit_HS}
 \intertext{for very thin channels and}
 \bracket{\frac{f}{\Delta\mathcal{P}}}_\mathrm{cr}=&\,\frac{1}{12}\,\frac{\av{W(x)^{-1}}_x}{\av{W(x)^{-3}}_x}=\,\frac{2\,\Delta\Omega^2\,\delta^2}{3\,\bracket{3+2\delta+3\delta^2}}, \label{eq:f_dp_crit_2D}
\end{align}
for infinite height channels $\Delta H\to \infty$.

\begin{figure}
\sidecaption
\resizebox{0.6\hsize}{!}{\includegraphics*{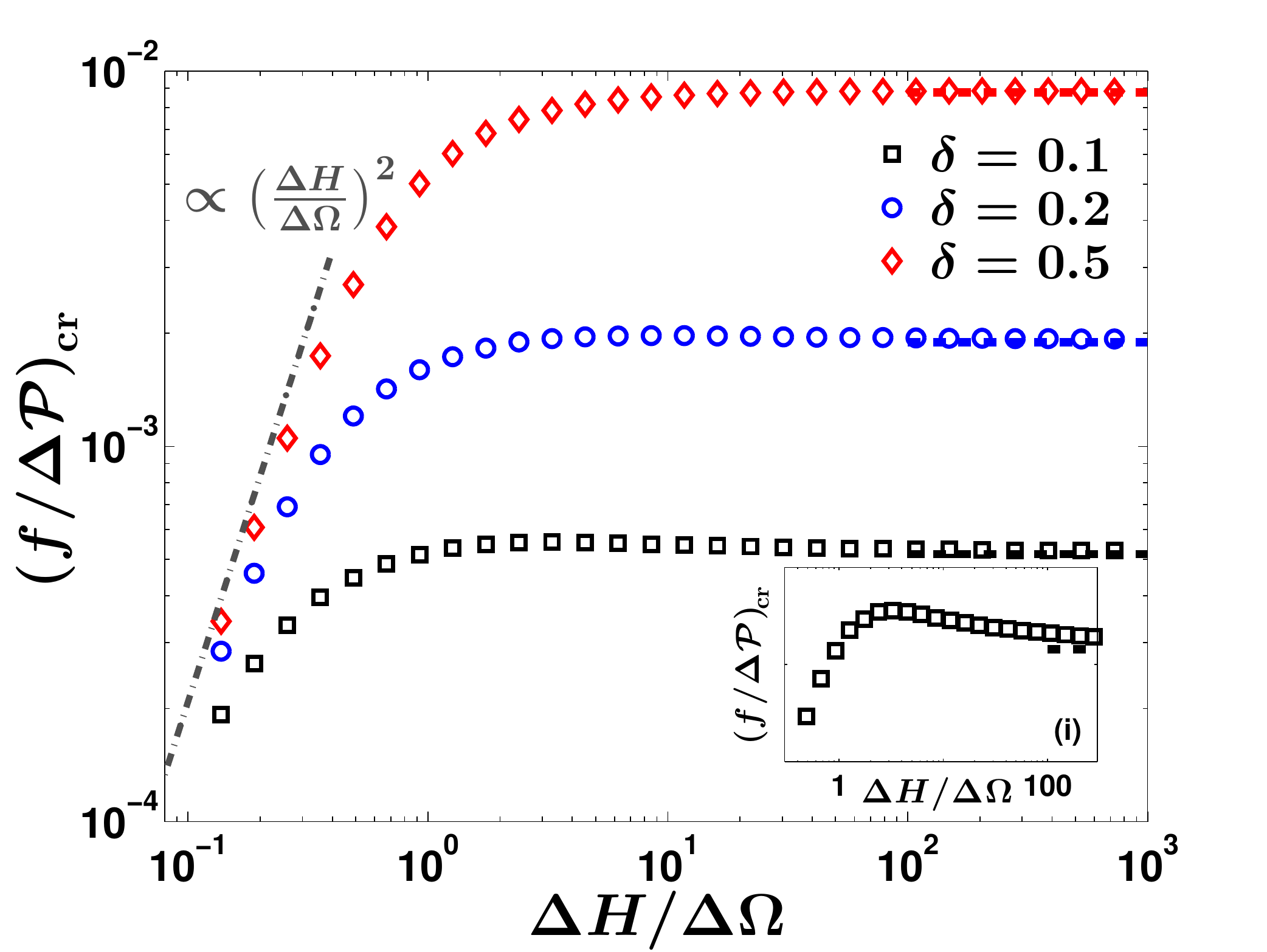}}
\caption{The dependence of the critical ratio $\bracket{f/\Delta\mathcal{P}}_\mathrm{cr}$
  on the channel height $\Delta H$ is depicted. The dashed dotted line represents the behavior for Hele-Shaw like channels, cf. \eq{eq:f_dp_crit_HS}, while the horizontal dashed lines indicate the asymptotic value for infinite height channels, \eq{eq:f_dp_crit_2D}. Inset: At an intermediate range of channels heights the critical ratio attains its maximum value. The maximal channel width is set to $\Delta \Omega=0.5$.}
  \label{fig:Fig4}
 \end{figure}

In \figref{fig:Fig4}, we depict the numerically evaluated results for $\bracket{f/\Delta \mathcal{P}}_\mathrm{cr}$ as a function of the channel height $\Delta H$ for various aspect ratios $\delta$. The critical ratio exhibits a non-linear dependence on the ratio of channel height to maximum width $\Delta\Omega$. While $\bracket{f/\Delta \mathcal{P}}_\mathrm{cr}$ grows independently of the particular local width $W(x)$ with $\Delta H^2$ for \textit{Hele-Shaw} channels it saturates at an asymptotic value for $\Delta H \gtrsim \Delta \Omega$. According to \eq{eq:f_dp_crit_2D}, these asymptotics are solely determined by the channel geometry. Espe\-cially, the smaller the maximum channel width $\Delta\Omega$ the less is the ratio $f/\Delta\mathcal{P}$ in order to inhibit
particle transport for a given aspect ratio $\delta$. While $\bracket{f/\Delta\mathcal{P}}_\mathrm{crit}$ goes to zero for almost closed channels $\delta \to 0$, \eq{eq:f_dp_crit_2D} resembles the Poiseuille flow result $(f/\Delta\mathcal{P})_\mathrm{cr}=\Delta \Omega^2/12$ for straight channels $\delta=1$. In between the two limits, viz., $\Delta H \to 0$ and $\Delta H \to \infty$, $\bracket{f/\Delta \mathcal{P}}_\mathrm{cr}$ attains a maximum for sufficient strong corrugated channels, i.e., $\delta \ll 1$. This fact is presented in depth in the inset (i) of \figref{fig:Fig4}. In particular, it turns out that channels with square cross-section $\Delta H \simeq \Delta\Omega$ can be treated as a quasi $2$D geometry and thus the curves for the mean particle velocity overlap for $\Delta H \gtrsim \Delta \Omega$, cf. \figref{fig:Fig3}(a).

The described non-linear behavior of the mean particle velocity $\av{\dot{x}}$ comes along with a peculiar behavior of the effective diffusion coefficient $D_\mathrm{eff}$, see \figref{fig:Fig3}(b). In particular, the variations in the flow field at the microscale contribute significantly to the macroscale effective diffusion coefficient $D_\mathrm{eff}$: For $\Delta\mathcal{P} < 1$, $D_\mathrm{eff}$ is mainly determined by the channel's geometry and the constant bias $f$; $D_\mathrm{eff}$ exhibits the known bell shaped behavior as a function of $f$ \cite{Burada2008}. At $\Delta\mathcal{P}\approx \Delta\mathcal{P}_\mathrm{cr}$, a drastic reduction of the diffusivity can be observed.
This effect, which occurs when the constant bias and the flow start to counteract such that the field $\V{F}(\V{q})=f \V{e}_x + \V{u}(\V{q})$ contains vortices
and stagnation points at given transverse height locations $z$, see \figref{fig:Fig1}, is referred to as \textit{hydrodynamically enforced entropic trapping} (HEET).
With the further growth in $\Delta\mathcal{P}$, $D_\mathrm{eff}$ exhibits Taylor-Aris dispersion \cite{Taylor1953,Aris1956} irrespective of the channel constriction, i.e., $D_\mathrm{eff}\propto \Delta\mathcal{P}^2$.

Obviously, the behavior of $D_\mathrm{eff}$ on $\Delta\mathcal{P}$ depends crucial on the channel height. For \textit{Hele-Shaw} like channels, $\Delta H \ll \Delta\Omega$, we observe a similar dependence of $D_\mathrm{eff}$ on $\Delta \mathcal{P}$ known from our previous $2$D studies \cite{Martens2013,Martens2013b}. There, the effective diffusion coefficient is of the order of $D_0$ for
$\Delta \mathcal{P} \lesssim \Delta \mathcal{P}_\mathrm{cr}$ and displays an abrupt decrease at $\Delta \mathcal{P}_\mathrm{cr}$. Remarkable, the values of $D_\mathrm{eff}$ are several orders of magnitudes smaller than the bulk value. This HEET-effect becomes more pronounced for thinner channels ($4$ orders of magnitude for $\Delta H/\Delta \Omega=0.1$) and larger external bias $f$ (not shown explicitly).

\begin{figure}
 \includegraphics[width=0.49\linewidth]{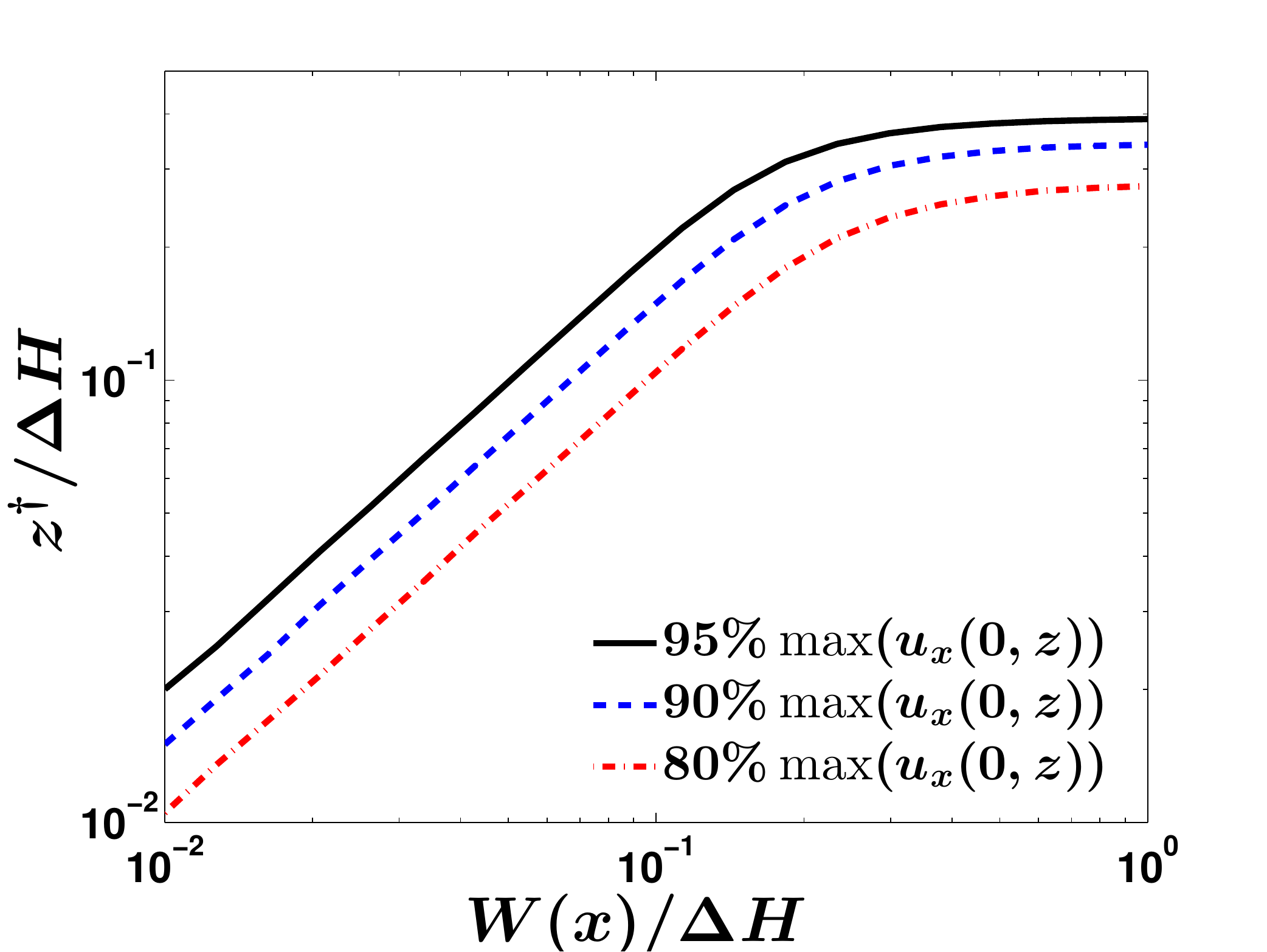}
 \includegraphics[width=0.49\linewidth]{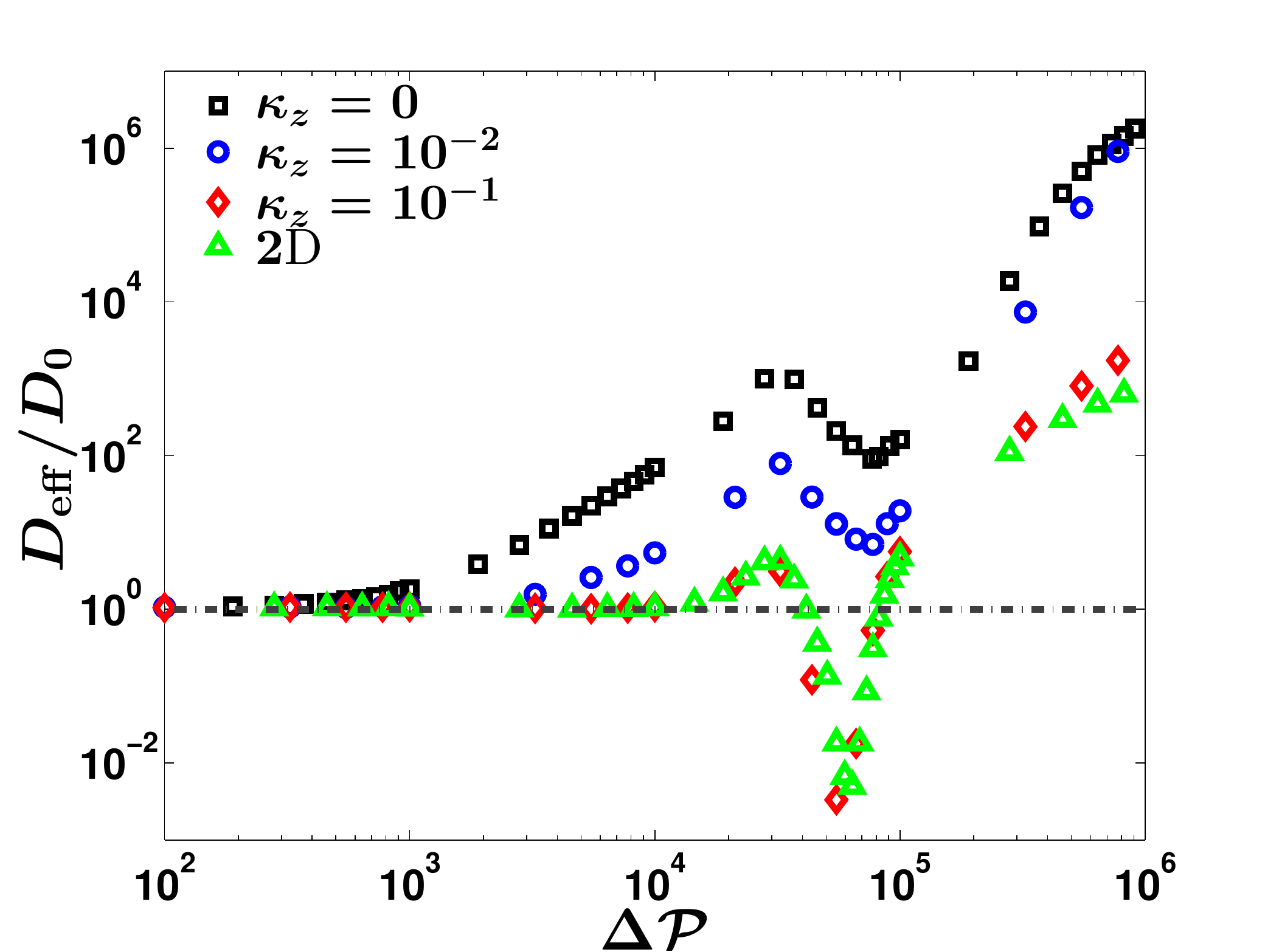}
 \caption{\label{fig:Fig5} Left panel: Critical transverse position $z^\dagger$, where the flow velocity attains $80\%$ (dash-dotted line), $90\%$ (dashed line), or $95\%$ (solid line) of the maximum flow velocity, as a function of the local channel width $W(x)$. For high channels $W(x)\ll \Delta H$, the critical transverse position is proportional to the local width, $z^\dagger \propto W(x)$. Right panel: Impact of confining potential $\Phi_0^ z(z)= 0.5 \kappa_z (z-\Delta H/2)^2$ on the effective diffusion coefficient. Superimposed are the numerical results (triangles) for the purely $2$D setup \cite{Martens2013}. The remaining parameter values are $\Delta\Omega=0.5$, $\Delta\omega=0.1$, $\Delta H=50$, and $f=100$.}
\end{figure}

Contrary to our reasoning in the previous $2$D studies \cite{Martens2013,Martens2013b}, $D_\mathrm{eff}$ does not depict the typical HEET characteristics for the quasi $2$D setup, $\Delta H \gg \Delta\Omega$. Nevertheless, given that the value of $D_\mathrm{eff}$ attains its minimum at $\Delta\mathcal{P}_\mathrm{cr}$, the effective diffusion coefficient grows proportional to $\Delta\mathcal{P}^2$ for intermediate to large pressure drops $\Delta \mathcal{P}$. As mentioned earlier, the velocity profile for $u^x$ is flat in $z$ except near the walls, cf. square markers in \figref{fig:Fig2}. FEM simulation results for the Stokes equation in a $3$D channel, \eq{eq:SE}, indicate that $u^x$ grows linearly in $z$ for $0\leq z \lesssim W(x)$, respectively, decreases linearly $\Delta H - W(x) \lesssim z \leq \Delta H$, see left panel in \figref{fig:Fig5}. In between, the flow profile is flat in $z$-direction and depends solely on $x$ and $y$. In this central region, the local force field $\V{F}(\V{r})=\,f\V{e}_x+\V{u}
$ contains vortices pushing the particles towards the channel wall. Therefore, the particles exhibit long residence times in the domains of strong accumulation although they experience continuous thermal fluctuations. This is the microscopic origin of the \textit{hydrodynamically enforced entropic trapping} effect. In the absence of an additional scalar potential $\Phi^ z(z)$ confining the particles in $z$-direction, the particles are equally distributed in $z$-direction. Consequently, particles located close to the plane walls at $z=0,\Delta H$ are dragged by the external bias $f$ and move to the right. Simultaneously, particles in the central layer may move to the left or slowly to the right or are even be trapped. This leads to the observed drastic enhancement of diffusion $D_\mathrm{eff} \propto \Delta P^2$.

In the right panel of \figref{fig:Fig5}, we present the impact of an additional confining scalar potential
\begin{align}
\Phi^ z(z)= \frac{\kappa_z}{2} \bracket{z-\frac{\Delta H}{2}}^2,
\end{align}
on the effective diffusion coefficient $D_\mathrm{eff}$. With growing trap stiffness $\kappa_z$ the particle becomes confined around the central layer at $z=\Delta H/2$. Consequently, the impact of the plane walls at $z=0,\Delta H$ diminishes and, as expected, the behavior of $D_\mathrm{eff}$ on $\Delta \mathcal{P}$ resembles the numerical results (triangles) for the purely $2$D setup \cite{Martens2013}.

\section{\label{sec:conclusion} Conclusions}

Bearing microfluidic application in mind, we addressed here the problem of transport of Brownian particles in a $3$D channel geometry confined by sinusoidally modulated walls and two plane walls placed at $z=0$ and $z=\Delta H$. The analysis of particle transport caused by the counteraction of a pressure-driven flow
and a constant bias of strength $f$ pointing in the opposite direction, ensues the intriguing finding that the \textit{hydrodynamically enforced entropic trapping} effect \cite{Martens2013,Martens2013b} crucially depends on the channel height $\Delta H$. In particular, the flow profile in channel direction changes from parabolic shape $u^x \propto z\bracket{\Delta H -z}$ for very thin channels -- \textit{Hele-Shaw} limit $\Delta H \ll \Delta \Omega$ -- to an almost flat profile in $z$ (except near the walls) in the limit of infinite extended channels $\Delta H \gg \Delta \Omega$. Therefore, we found that the critical ratio of force magnitude $f$ to the applied pressure drop $\Delta \mathcal{P}$ at which the mean particle current vanishes identically and the effective diffusion coefficient is significantly suppressed -- HEET effect -- scales with $\Delta H^2$ for thin channels and saturates at an asymptotic value for high channels.
Consequently, the critical pressure drop increases with shrinking channel width $\Delta \mathcal{P}_\mathrm{cr} \propto f_\mathrm{cr}/\Delta H^2$ leading to a giant suppression of the effective diffusivity; 4 orders of magnitude compared to the bulk value for the parameters explored in this work. We emphasize that the HEET effect becomes more pronounced for larger $f_\mathrm{cr}$ and $\Delta\mathcal{P}_\mathrm{cr}$, respectively, leading to stiffer traps \cite{Martens2013}. Consequently, we expect that even smaller values of $D_\mathrm{eff}/D^0$ might be observed for stronger forces $f$. The latter feature may prove advantageous in tailoring more efficient separation devices as compared to currently used methods in separating micro- and/or nano-sized particles. Note that our methodology admits the situation of a pressure-driven solvent flow; alternatively, similar effects can be expected in electroosmotic and electrokinetic systems or in a resting solvent with nonvanishing divergence-free forces.

Finally, we briefly mention the influence of hydrodynamic interaction between the particles and the corrugated boundaries. In general, the
interaction of a particle with the wall depends on the particle shape \cite{Zoettl2013}, orientation, and position, as well as on the geometry of the channel walls \cite{Happel1965,DiCarlo2007}. Generally, the presence of a flat wall is known to effectively retard the motion of a small spherical particle. This effect can be roughly accounted for by introducing an additional drag force with magnitude
$\Delta f=\,R\,f/(l-R)+\mathcal{O}\bracket{(R/l)^2}$ \cite{Happel1965}, acting against the direction of the particle's motion. Here, $R$ is the radius of particle, and $l$ is the distance to the wall. For two planar parallel confining walls, this force becomes more complicated, see, e.g., \cite{Liron1976}. If we consider a deviation of $10\%$, $\Delta f = 0.1 f$, the distance between particle and wall has to be at least $l\geq 10 R$ to formally disregard the impact of the hydrodynamic coupling to the wall. For this reason, the theory is formally justified for a dilute particle suspension. Otherwise, the impact of hydrodynamic interactions is expected to originate and become significant at the bottleneck. For instance, accumulation of particles at the constricting bottleneck may enhance due to the interaction with the wall \cite{Schindler2007,Reddig2013}.

\clearpage

\appendix

\section{\label{app:genFJ} Derivation of the Fick-Jacobs solution for generalized forces}

For finite corrugation $\e \neq 0$, it is feasible to measure the transverse
coordinate $y\to \e\,y$ and the boundary functions $\omega_\pm(x)\to \e\,h_\pm(x)$ in units of $\e$.
Consequently, the gradient $\grad\to(\partial_x,\e^{-1}\partial_y,\partial_z)^T$ and the Laplace
operator $\bigtriangleup \to (\partial_x^2+\e^{-2}\partial_y^2+\partial_z^2)$
change. Due to the invariance of $\grad \times \grb{\Psi}$, the vector potential scales as
$\grb{\Psi}\to \bracket{\e\,\Psi_x,\Psi_y,\e\,\Psi_z}^T$. Further, we expand the probability density function (PDF)
$P(\V{q},t)=P_0(\V{q},t)+\e^2\,P_1(\V{q},t)+\mathcal{O}(\e^4)$ as well as
$\Phi(\V{q})=\Phi_0(\V{q})+\e^2\,\Phi_1(\V{q})+\mathcal{O}(\e^4)$, and each component of
$\Psi(\V{q})_i=\Psi_i^0(\V{q})+\e^2\,\Psi_i^1(\V{q})+\mathcal{O}(\e^4)$, for $i=x,y,z$, in
the a series in even orders of $\e$. Substituting this ansatz into
\eq{eq:smol}, yields
\begin{align}
 0=&-\partial_y\left[e^{-\Phi_0}\partial_y\bracket{e^{\Phi_0}\,P_0}\right]
+\e^2\left\{\partial_tP_0 -
\partial_y\left[e^{-\Phi_1}\partial_y\bracket{e^{\Phi_1}\,P_0}
\right]-\partial_y\left[e^{-\Phi_0}\partial_y\bracket{e^{\Phi_0}\,P_1}\right]
\right. \nonumber \\ &\left.
-\partial_x\left[e^{-\Phi_0}\partial_x\!\bracket{e^{\Phi_0}\,P_0}\right]
-\partial_z\left[e^{-\Phi_0}\partial_z\!\bracket{e^{\Phi_0}\,P_0}\right]+\bracket{\grad \times\grb{\Psi}_0}\cdot\grad P_0\right\}+\mathcal{O}(\e^4).
\label{eq:appB_pdf_series}
\end{align}
Furthermore, the no-flux boundary conditions read:
\begin{subequations}
 \begin{align}
 &\mbox{at}\,y=h_+,h_-: \nonumber \\ &0=\mp
e^{-\Phi_0}\partial_y\bracket{e^{\Phi_0}\,P_0}\mp\,\e^2\left\{e^{-\Phi_0 }
\partial_y\bracket{e^{\Phi_0}\,P_1}+e^{-\Phi_1}\partial_y\bracket{e^{\Phi_1}\,
P_0}\right. \nonumber\\
&\left.-h_\pm'(x)e^{-\Phi_0}\partial_x\bracket{e^{\Phi_0}\,P_0}
+h_\pm'(x)\bracket{\grad\times\grb{\Psi}_0}_x\,P_0-\bracket{\grad\times\grb{\Psi
}_0}_y\,P_0\right\}+\mathcal{O}(\e^4),\\
 &\mbox{at}\,z=\Delta H,0: \nonumber \\
 &0=\mp
e^{-\Phi_0}\partial_z\bracket{e^{\Phi_0}\,P_0}\pm
\bracket{\grad\times\grb{\Psi}_0}_z\,P_0\mp\e^2\left\{e^{-\Phi_0}
\partial_z\bracket{e^{\Phi_0}\,P_1}\right. \nonumber\\
&\left.+e^{-\Phi_1}\partial_z\bracket{e^{ \Phi_1
}\,P_0}-\bracket{\grad\times\grb{\Psi}_1}_z\,P_0-\bracket{\grad\times\grb{\Psi}
_0}_z\,P_1\right\}+\mathcal{O}(\e^4). \label{eq:appB_bc_top}
\end{align}\label{eq:appB_bc_series}
\end{subequations}
\noindent From the leading order $\e^0$, Eqs.~\eqref{eq:appB_pdf_series} and
\eqref{eq:appB_bc_series},  immediately follows that
$P_0(\V{q},t)=g(x,t)e^{-\Phi_0(\V{q})}$, where $g(x,t)$ is an unknown
function which has to be determined from the second order, $\mathcal{O}(\e^2)$, of
\eq{eq:appB_pdf_series}. Thereby, we assume that the translation invariance in the $z$-direction is broken solely by the scalar potential $\Phi$. Note that this assumption is always justified for laminar flows in confinements, because $u^z=\bracket{\grad\times\grb{\Psi}}_z=0$. Otherwise, the unknown function $g$ can also be dependent on $z$.

Integrating $\mathcal{O}(\e^2)$ over the local cross-section
$Q(x)=\Delta H\bracket{h_+(x)-h_-(x)}$, rearranging derivatives and integrals,
and taking account of the no-flux boundary conditions, \eqs{eq:appB_bc_top}, finally, we arrive at
\begin{align}
  \partial_t P_0(x,t)=\! -\partial_xJ_0^x(x,t) =\! -\partial_x\!\!\intl{h_-(x)}{h_+(x)}{y}\!\!\intl{0}{\Delta
H}{z}\!\left[-e^{-\Phi_0}\partial_x g+\bracket{\grad\times\grb{\Psi}_0}
_x g\, e^{-\Phi_0} \right] \label{eq:appB_pdf_int}
\end{align}
with the marginal probability flux $J_0^x(x,t)$.

In what follows, we focus on the steady state solution, i.e., $\partial_t
P_0=0$. Since the unknown function $g$ is assumed to be independent of
$z$, \eq{eq:appB_pdf_int} simplifies to
\begin{align}
const = J_0^x=&\,-e^{-A(x)}\,g'(x)+g(x)\,e^{-A(x)}\,\alpha(x).
\end{align}
Thereby, the ``entropic'' potential $A(x)$ and the accessory part $\alpha(x)$ read
\begin{align}
A(x)=&\,-\ln\left[\intl{h_-(x)}{h_+(x)}{y}\intl{0}{\Delta H}{z}e^{-\Phi_0}\right],\\
\intertext{and}
 \alpha(x)=&\,\intl{h_-(x)}{h_+(x)}{y}\intl{0}{\Delta H}{z}\,\bracket{\grad\times\grb{\Psi}_0}
_x\,e^{-\Phi_0(\V{q})+A(x)},
\end{align}
respectively. Solving the differential equation for $g(x)$ yields
\begin{align} \label{eq:appB_solpdf}
 P_0(\V{q})=\,g(x)\,e^{-\Phi_0(\V{q})}=\,\left[C_0-J_0^x\,\intl{0}{x}{x'}\,e^{\mathcal{F}(x')}\right]\,e^{
-\Phi_0(\V{q})+\intl{0}{x}{x'}\alpha(x')},
\end{align}
where $C_0$ is an integration constant and $\mathcal{F}(x)$ is the potential of mean force in the longitudinal direction,
\begin{align}
 \mathcal{F}(x)\!=\!-\ln\left[\intl{h_-(x)}{h_+(x)}{y}\!\!\intl{0}{\Delta H}{z}\!
e^{-\Phi_0}\!\right]\! -\!\intl{0}{x}{x'}\!\!\intl{h_-(x')}{h_+(x')}{y}\!\!\intl{0}{\Delta H}{z}\!
\bracket{\grad \times \grb{\Psi_0}}_x\!P_\mathrm{eq}(y,z|x').
\end{align}
Here, $P_\mathrm{eq}(y,z|x)$ represents the equilibrium PDF of $y$ and $z$,
conditioned on $x$, given by $P_\mathrm{eq}(y,z|x)=\exp[-\Phi_0(\V{q})+A(x)]$.

Plugging the solution for $g(x)$, \eq{eq:appB_solpdf}, into \eq{eq:appB_pdf_int}, one obtains an ordinary differential equation for the stationary marginal PDF $P_0(x)$
\begin{align}
 0= \, \partial_x \left[\pt{\mathcal{F}}{x}P_0(x)\right]+\partial_x^2 P_0(x).
\end{align}
Moreover, the kinetic equation for the time-dependent marginal PDF $P_0(x,t)$ can be
drawn from its steady state solution, resulting in the \textit{generalized Fick-Jacobs equation}
\begin{align} \label{eq:appB_genFJ}
 \partial_t P_0(x,t)=\,\partial_x\left[\pt{\mathcal{F}(x)}{x} P_0(x,t)\right]
+\partial_x^2 P_0(x,t).
\end{align}
We emphasize that the generalized Fick-Jacobs equation \eq{eq:appB_genFJ} can also be derived from \eq{eq:appB_pdf_int} assuming separation of time scales \cite{Zwanzig1992,Reguera2001,Burada2008}. By using $P_0(\V{q},t)=g(x,t)e^{-\Phi_0}$, \eq{eq:appB_pdf_int} can be written as
\begin{align} \label{eq:appB_pdf_int2}
  \partial_t P_0(x,t)=-\partial_x\intl{h_-(x)}{h_+(x)}{y}\intl{0}{\Delta
H}{z}\,\left[\bracket{-\partial_x \Phi_0 + \bracket{\grad\times\grb{\Psi}_0}
_x} P_0(\V{q},t) - \partial_x P_0(\V{q},t)\right].
\end{align}
This relation connects the marginal PDF $P_0(x,t)$ with the $P_0(\V{q},t)$. According to the Bayes theorem, the PDF is given by the
product of the conditional PDF $P(y,z|x,t)$ and the marginal PDF, viz. $P(\V{q},t)=\,P(y,z|x,t)\,P(x,t)$.  In general, for arbitrary channels and forces, $P(y,z|x,t)$ cannot be calculated analytically. In the presence of time scales separation, however, it is reasonable to assume that the distribution of the transverse coordinates $y$ and $z$ relaxes much faster to the equilibrium one than that of the longitudinal coordinate. Hence, the conditional PDF can be approximated by its equilibrium solution $P(y,z|x,t)\to\,P_\mathrm{eq}(y,z|x)=\exp[-\Phi_0(\V{q})+A(x)]$. Then, it becomes possible to show that \eq{eq:appB_pdf_int2} transforms into the generalized Fick-Jacobs equation \eq{eq:appB_genFJ}.

Note that although an equation for the marginal PDF $P_0(x,t)$ or, respectively, for the stationary PDF $P_0(\V{q})$, can be derived, a closed analytic solution obeying the periodicity requirement $P(x+1,y,z,t)=P(x,y,z,t)$ can only be obtained if (i) the curl-free potential scales $\Phi(\V{q})\sim-f\,x^\beta$ with $\beta=0,1$ and (ii)
$(\grad\times\grb{\Psi})_x$ is periodic in $x$. In order to calculate the stationary periodic solution, the integration constant $C_0$ in \eq{eq:appB_solpdf} has to be fixed
\begin{align*}
 P_0(x+1,y,z)\!=&\!\left[C_0-J_0^x\intl{0}{1}{x} e^{\mathcal{F}(x)}- J_0^x \intl{0}{x}
{x'} e^{\mathcal{F}(x')}\right]\!e^{-\,\Delta\mathcal{F}} e^{-\Phi_0(\V{q})+\intl{0}{x}{x'}\alpha(x')} \\ \equiv&\,P_0(x,y,z).
\end{align*}
In the case that $\mathcal{F}$ is periodic in $x$, i.e., $\Delta\mathcal{F}=\mathcal{F}(x+1)-\mathcal{F}(x)=0$, the solution to
\eq{eq:appB_solpdf} fulfills the periodicity requirement provided that either $\int_{0}^{1}\mathrm{d}x\, \exp(\mathcal{F}(x))=0$ or the probability current vanishes, $J_0^x=0$. The former
condition is only feasible for $\mathcal{F}(x)=-\infty$, which is 
unphysical. From the latter condition, $J_0^x=0$, follows immediately
that divergence-free force has to vanish for all values of $x$, i.e., $(\grad\times\grb{\Psi})_x=0$. Consequently, $\Phi(x+1,y,z)-\Phi(x,y,z)$ must vanish identically and
therefore the stationary PDF is constant, $P_0(\V{q})=C_0$. Then, the marginal
PDF scales with the local channel cross-section
\begin{align}
 P_0(x)\sim\,\Delta H\bracket{h_+(x)-h_-(x)}.
\end{align}
In the opposite limit, $\Delta\mathcal{F}\neq0$, the periodicity requirement
is solely fulfilled for $C_0=J_0^x\int_{0}^{1}\mathrm{d}x\,\exp(\mathcal{F}(x))\,/(1-\exp(\Delta\mathcal{F}))$ and finally we end up with
\begin{align}
 P_0(x)=\,\frac{e^{-\mathcal{F}(x)}\,\intl{x}{x+1}x'\,e^{\mathcal{F}
(x')}}{\intl{0}{1}x\,e^{\mathcal{F}
(x)}\intl{x}{x+1}x'\,e^{\mathcal{F}
(x')}},
\end{align}
where the stationary PDF obeys the normalization condition
$\int_{\mathrm{unit-cell}} P(\V{q},t)\,\mathrm{d}^3\V{q} =1$. We stress that if $\beta>1$ or the
longitudinal coordinate $x$ is multiplicatively connected to the transverse
coordinates, a closed periodic solution for $P_0(x)$ cannot be found \cite{Haenggi1982,Risken}.

\begin{acknowledgement}
This work has been supported by the Deutsche Forschungsgemeinschaft via SFB 910 (S.~M.) and IRTG 1740 (A.~V.~S. and L.~S.-G.).
\end{acknowledgement}

\providecommand*{\mcitethebibliography}{\thebibliography}
\csname @ifundefined\endcsname{endmcitethebibliography}
{\let\endmcitethebibliography\endthebibliography}{}

\end{document}